\documentstyle{mn}

\title{RX J2115--5840: confirmation of a new near-synchronous polar}
\author[G. Ramsay et al]
{Gavin Ramsay$^{1}$, David A. H. Buckley$^{2}$, Mark Cropper$^{1}$,
M. K. Harrop-Allin$^{1}$\\
$^{1}$Mullard Space Science Laboratory, University College London,
Holmbury St.Mary, Dorking, Surrey, RH5 6NT\\
$^{2}$South African Astronomical Observatory, PO Box 9, Observatory 7935, Cape
Town, South Africa\\
}

\date{accepted: 21 Oct 1998}

\begin{document}

\maketitle

\begin{abstract} 

Following the suggestion of Schwope et al that the magnetic
cataclysmic variable RX J2115--5840 maybe a near-synchronous polar, we
obtained optical polarimetry of this system over a 2 week period. From
a power spectrum of the circular polarimetry data we determine that
the spin period of the white dwarf and the binary orbital period which
differ by 1.2 percent. RX J2115--5840 is thus the fourth near
synchronous polar and has the shortest spin-orbit beat period: 6.3
days. By folding the data on spin, beat and orbital periods we find
evidence that the accretion stream is directed towards opposite
magnetic poles as the stream precesses around the white dwarf on the
spin-orbit beat period. The phasing requires that the accretion flow
must be directed onto the same magnetic field line at all spin-orbit
beat phases implying that at some phases the flow must follow a path
around the white dwarf before accreting. This is difficult to
reconcile with simple views of how the accretion stream attaches onto
the magnetic field of the white dwarf.

\end{abstract}


\vspace{2.5cm}

\section{Introduction}

EUVE 2115--58.6 was discovered during the {\sl EUVE} all sky survey
(Bowyer et al 1996) and also during the {\sl ROSAT} all sky survey (RX
J2115--5840, Voges et al 1997). From optical spectra, Craig (1996)
suggested that RX J2115--5840 is a magnetic Cataclysmic Variable
(mCV). Further spectroscopic observations by Vennes et al (1996)
suggested an orbital period ($P_{o}$) of 110.8 min or an alias at
102.8 min.  Optical polarimetry obtained by Schwope et al (1997)
showed variable circular polarisation of up to 15 percent, indicating
that it is a member of the polar sub-class of mCVs. In these systems
the accreting white dwarf has a sufficiently strong magnetic field to
lock the spin of the white dwarf into synchronous rotation with the
binary period. The bulk of the accretion luminosity is liberated at
X-ray/EUV wavelengths.

Schwope et al (1997) also suggested that the orbital period differs
with respect to the spin period of the white dwarf by $\sim1\%$. They
concluded that the spectroscopic period of 110.8 min represents the
binary orbital period, while the photometric period of 109.84 min or
109.65 min represents the spin period of the white dwarf ($P_{s}$). If
confirmed, this would make RX J2115--5840 the fourth near-synchronous
polar and the first one below the 2--3 hr orbital period
gap. Therefore to determine if RX J2115--5840 is indeed
near-synchronous, we obtained white light polarimetric observations
covering 2 weeks.

\section{Observations}

RX J2115--5840 was observed in white light using the SAAO 1.9m
telescope and UCT Polarimeter (Cropper 1985) between 1997 July 29 and
1997 August 11. Data were obtained on 10 nights over this interval
although the total amount of data obtained on each night varied.
Conditions ranged from photometric to non-photometric. Sky
measurements were obtained every 15--25 mins and subtracted from the
source measurements by a polynomial fit to the sky data. Polarised and
non-polarised standard stars (Hsu \& Breger 1982) and calibration
polaroids were observed at the beginning of the night to set the
position angle offsets and efficiency factors.

\section{Results}

\begin{figure}
\begin{center}
\setlength{\unitlength}{1cm}
\begin{picture}(8,12.5)
\put(-1.5,-1.5){\includegraphics{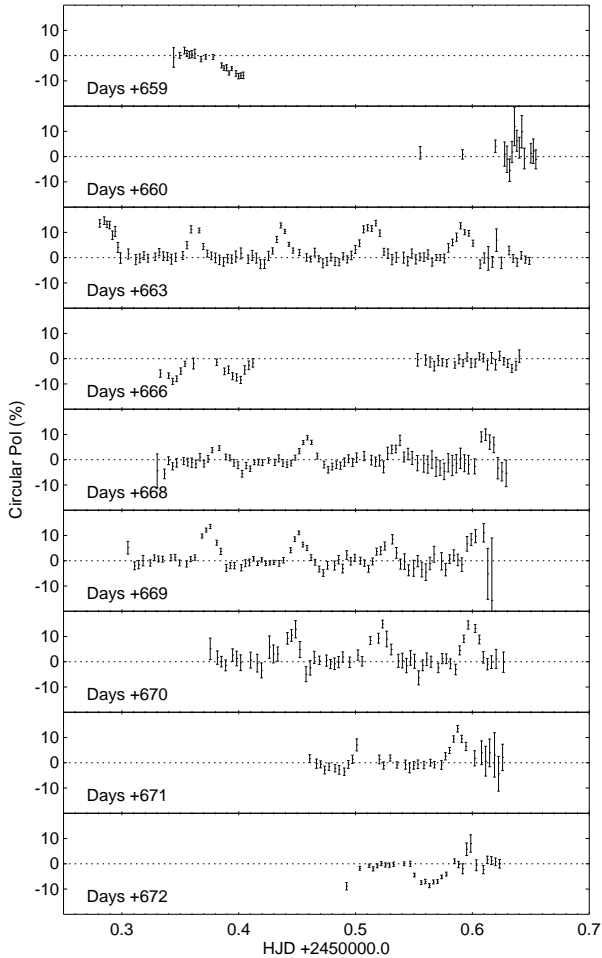}}
\end{picture}
\end{center}
\caption{The circular polarisation data. The label in the left hand
of each panel indicates the HJD +2450000.0.}
\label{cpol} 
\end{figure}

The white light intensity data are similar to those shown in Schwope
et al (1997) in that unlike most polars they do not show large
photometric variations, although humps were seen in some of the data
sets. We recorded low levels (typically 3--4 percent) of linear
polarisation. In this paper, however, we concentrate on the circular
polarisation data since the intensity data were compromised to some
extent by non-photometric conditions on some nights. The circular
polarisations are shown in Fig.  \ref{cpol}. For most of the
observation the circular polarisation is either close to zero or shows
positive excursions which follow the intensity curve at some
epochs. However, there are occasions when negative circular
polarisation is seen (HJD 2450000+ 659, 666 and 672). Similar
behaviour was seen by Schwope et al (1997), although they had circular
polarisation from short sections of data on only 3 days.

In fully synchronised polars, the spin period of the white dwarf is
equal to the binary orbital period. In these systems the circular
polarisation curves are broadly repeatable from cycle to cycle since
the accretion flow is directed onto the same accretion pole(s). These
observations showing modulated circular polarisation of positive
polarity at some epochs and negative polarity at other epochs suggests
that RX J2115--5840 is not fully synchronised.

\begin{figure*}
\begin{center}
\setlength{\unitlength}{1cm}
\begin{picture}(12,10)
\put(-3,-28.2){\includegraphics{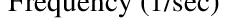}}
\end{picture}
\end{center}
\caption{From the top: the amplitude spectrum of the circular polarisation
data; the amplitude spectrum in greater
detail over three frequency intervals; the window function and the 
pre-whitened spectrum (see text for details).}
\label{dft} 
\end{figure*}

While a Discrete Fourier Transform (DFT) may not be the optimum method
to analyse intensity or polarisation curve of a near-synchronous polar
since there may be phase shifts between the curves for each pole, we
use a DFT for this first stage of our investigation. Fig. \ref{dft}
shows the amplitude spectrum of the circular polarisation data without
subtracting the mean. The highest amplitude peak corresponds to a
period of 110.889 mins -- similar to the spectroscopic period of 110.8
mins reported by Vennes et al (1996). Moving higher in frequency, the
next peak in amplitude corresponds to a period of 109.547 mins --
similar to the shorter of the two possible optical photometric
periods, 109.84 and 109.65 mins, reported by Schwope et al (1997).

We take the periods 110.889 and 109.547 found from our DFT to
represent the binary orbital frequency, $\Omega$, and the spin
frequency of the white dwarf, $\omega$, respectively. A side band
frequency 2$\omega - \Omega$ corresponding to a period of 108.237 mins
is then very close to an amplitude peak at 108.38 mins in the DFT
(this side band frequency is prominent in the near-synchronous polar
BY Cam: Mason et al 1998). At much lower frequencies, the spin-orbit
beat frequency $\omega-\Omega$, which corresponds to a period of 6.28
days, is within the FWHM of the highest amplitude peak seen at the
lowest frequencies (7.05 days). The next two highest frequencies in
the DFT (1.19 and 0.88 days) are aliases of this spin-orbit beat
frequency.

So far we can therefore account for the principle amplitude peaks in the DFT at
frequencies lower than 0.0002 Hz. Moving to the first harmonics of the
proposed spin and orbital periods (close to 0.0003 Hz), we find that
twice our proposed spin frequency, 2$\omega$, corresponds exactly with
a prominent amplitude peak. The prominent peak at
3.02442$\times10^{-4}$ Hz (=55.107 mins) is the $\omega+\Omega$
side-band frequency. We also detect the 3$\omega$ harmonic and the
4$\Omega - 3\omega$ side-band frequency

To test our proposed values of $P_{o}$ = 110.889 mins and $P_{s}$ =
109.547 mins, we pre-whitened the circular polarisation data with
these two frequencies (and their second and third harmonics), the
2$\omega - \Omega$ side band frequency ($P=$108.38 mins), the
4$\Omega-3\omega$ side band frequency ($P$=114.956 mins) and the more
$\omega\pm\Omega$ side band frequencies (shown in the lower panel of
Fig. \ref{dft}). All frequencies which have an amplitude greater than
1 percent have been removed. This implies that if any other
frequencies are present in the amplitude spectrum then they are
present at a low level.  We show in Table \ref{periods} the peak
amplitudes of all the frequencies we can distinguish, together with
their equivalent periods. We note that not all of the frequencies
quoted in Table \ref{periods} are necessarily significant (it is
difficult to assign significance levels to amplitude peaks). We have
merely noted those combinations of spin and orbital frequencies which
coincide with a peak in the amplitude spectrum.

\begin{table}
\begin{tabular}{lrr}
\hline
Frequency& Hz& Period (min)\\
\hline
$\Omega$ &1.502993$\times10^{-4}$&110.889\\
3$\Omega$&4.543406$\times10^{-4}$&36.683\\
$\omega$&1.521423$\times10^{-4}$&109.547\\
2$\omega$&3.041679$\times10^{-4}$&54.794\\
3$\omega$&4.559762$\times10^{-4}$&36.552\\
$\omega-\Omega$&1.6276$\times10^{-6}$&10238\\
$\omega$+$\Omega$&3.022470$\times10^{-4}$&55.143\\
2$\omega$+$\Omega$&1.539853$\times10^{-4}$&108.235\\
4$\Omega$-3$\omega$&1.449825$\times10^{-4}$&114.956\\
\hline
\end{tabular}
\caption{The frequencies in the amplitude spectrum which we can 
distinguish: we show the frequency in Herz and its equivalent period
in mins. We do not necessarily claim that all of these frequencies are
significant.}
\label{periods}
\end{table}

\section{The beat period}

\begin{figure}
\begin{center}
\setlength{\unitlength}{1cm}
\begin{picture}(12,9.5)
\put(-1,-3){\includegraphics{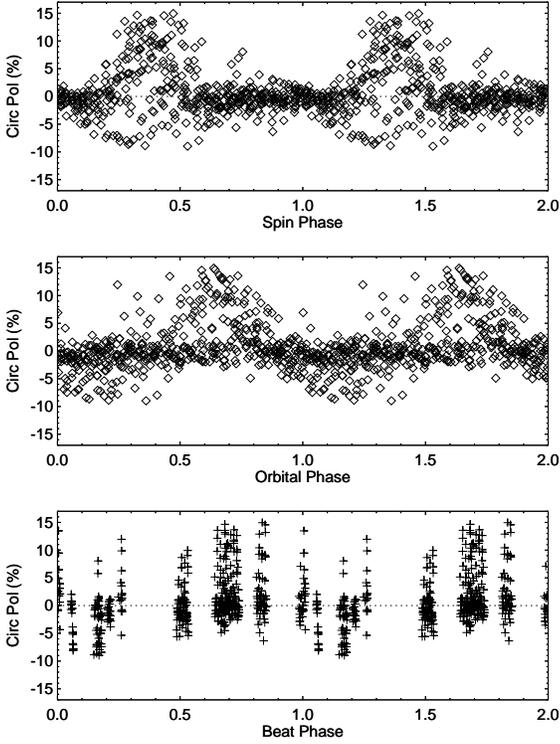}}
\end{picture}
\end{center}
\caption{The circular polarisation folded on the
spin period, 109.55 mins, (top panel); folded on the orbital period
(middle panel), folded on the orbital-spin
beat period, $\omega-\Omega$ = 6.28 days. Phase zero was arbitrarily
chosen to be HJD = 2450659.0.}
\label{beat_phase} 
\end{figure}

To make a more detailed investigation of the circular polarisation
data we folded the circular polarimetry on the proposed spin
period. The top panel of Fig. \ref{beat_phase} shows that for over
half the spin cycle the circular polarisation is close to zero, while
for the remainder of the cycle, there are positive or negative
excursions at approximately the same spin phase. If we fold the
circular polarimetry on the proposed orbital period (the middle panel
of Fig.  \ref{beat_phase}) we find similarly that the circular
polarisation is close to zero while there are phases of alternate
positive and negative circular polarisation.

To determine if these positive and negative excursions were observable
on the 6.28 day spin-orbit beat period we folded the circular
polarisation on this period. The bottom panel of Fig. \ref{beat_phase}
shows the folded data (phase zero was arbitrarily chosen to be
24500659.0, the start of our observation). Although there is a good
deal of variability due to spin variations at shorter time scales, the
mean level of circular polarisation is negative at
$\phi_{(\omega-\Omega)}\sim$0.07 and 0.17 while it increases at later
phases.
 
\begin{figure}
\begin{center}
\setlength{\unitlength}{1cm}
\begin{picture}(12,10)
\put(-1.5,-3){\includegraphics{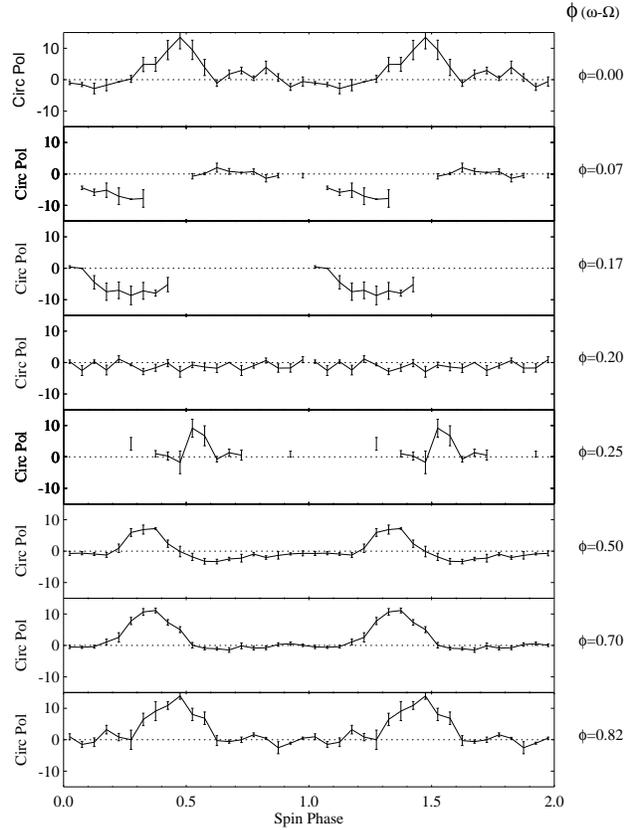}}
\end{picture}
\end{center}
\caption{The circular polarisation data folded and binned on the
proposed spin period of the white dwarf, 109.547 mins, as a function
of the spin-orbital beat period of 6.28 days. The beat phase is shown
at the right hand side.}
\label{beat_bin} 
\end{figure}

To examine variations in the circular polarisation data over the
spin-orbit beat period, we folded and binned each section of data
which corresponded to a discrete beat phase on the proposed spin
period of the white dwarf -- 109.547 mins (Fig. \ref{beat_bin}). As
expected from Fig. \ref{beat_phase}, at phases
$\phi_{(\omega-\Omega)}\sim$0.07 \& 0.17 the mean circular
polarisation is negative. The polarisation curve shows a negative
excursion lasting approximately half the spin cycle, while other spin
phases the polarisation is close to zero. At
$\phi_{(\omega-\Omega)}$=0.20 the polarisation is not significantly
modulated. At other beat phases a prominent positive hump is seen in
the folded spin polarisation curves, the peak of which advances in
phase as $\phi_{(\omega-\Omega)}$ increases.

\section{Pole-switching}

In polars the accretion flow leaves the secondary star on an initially
ballistic trajectory, couples onto the magnetic field of the white
dwarf primary and is directed below and/or above the orbital plane
before it is channeled onto the surface of the white dwarf. In fully
synchronous polars, the accretion flow is locked with respect to the
binary orbital rotation frame and the bulk of the accretion flow is
thought to be directed onto the prefered magnetic pole of the white
dwarf. However, in the case of near-synchronous polars, the accretion
flow rotates around the magnetic field of the white dwarf on the
timescale of the spin-orbit beat period. This has the effect that the
accretion flow will be directed preferentially onto first one then the
other magnetic pole of the white dwarf. At two phases of the
spin-orbit beat period we expect that the flow will be equally
directed onto both poles. This `pole-switch' will manifest itself most
obviously in the circular polarisation curves where the polarisation
will change sign after the accretion flow has `switched' poles. This
is seen in Fig. \ref{beat_bin} where at
$\phi_{(\omega-\Omega)}\sim$0.00 the polarisation is modulated with a
positive hump, but at $\phi_{(\omega-\Omega)}\sim$0.07 and 0.17 it is
modulated with a negative hump. Further, at
$\phi_{(\omega-\Omega)}\sim$0.20, the accretion flow is directed
equally towards both magnetic poles and the net polarisation is
zero. We expect that between $\phi_{(\omega-\Omega)}\sim$0.00 and 0.07
the net polarisation will also be zero. In a system where the spin
axis and the magnetic axes are both orthogonal to the binary plane
then it is possible that the accretion flow will be directed equally
onto both magnetic poles of the white dwarf and pole-switching will
not occur.

To make an estimate of the angle that the spin axis makes with the
magnetic axis ($m$) and the binary inclination ($i$) we examine the
predicted power spectra of the more rapidly rotating analogues of the
near-synchronous polars -- the intermediate polars (those mCVs where
typically $P_{s}=0.1P_{o}$). A number of authors have predicted the
power spectra of the light curves of IPs (Warner 1986, Wynn \& King
1992 and Norton, Beardmore \& Taylor 1996). The detection or
non-detection and relative strength of the individual components in
power spectra of IPs are dependent on a range of factors, for example
whether the system is discless or is stream-fed, whether the accretion
regions are directly opposite each other, $i$, $m$, the angle the
accretion regions subtend ($\beta$) around co-latitude $m$ and so
on. It is more difficult to invert this process and obtain values for
$i$, $m$ etc, from the power spectra: we can, however, make some
general remarks. Wynn \& King (1992) show that for low $i$ and low $m$
(where $i+m<90^{\circ}-\beta$ holds) only the spin-orbit frequency
$\omega-\Omega$ and its even harmonics will be detected. This is
clearly not the case in RX J2115--5840 and it is likely that this
system has high $i$ and high $m$ and obeys the condition
$i+m>90^{\circ}+\beta$. We should caution that in polars there are
less sites for reprocessing radiation compared to IPs since there is
no accretion disc.

\begin{figure*}
\begin{center}
\setlength{\unitlength}{1cm}
\begin{picture}(12,8)
\put(-1.5,12){\includegraphics{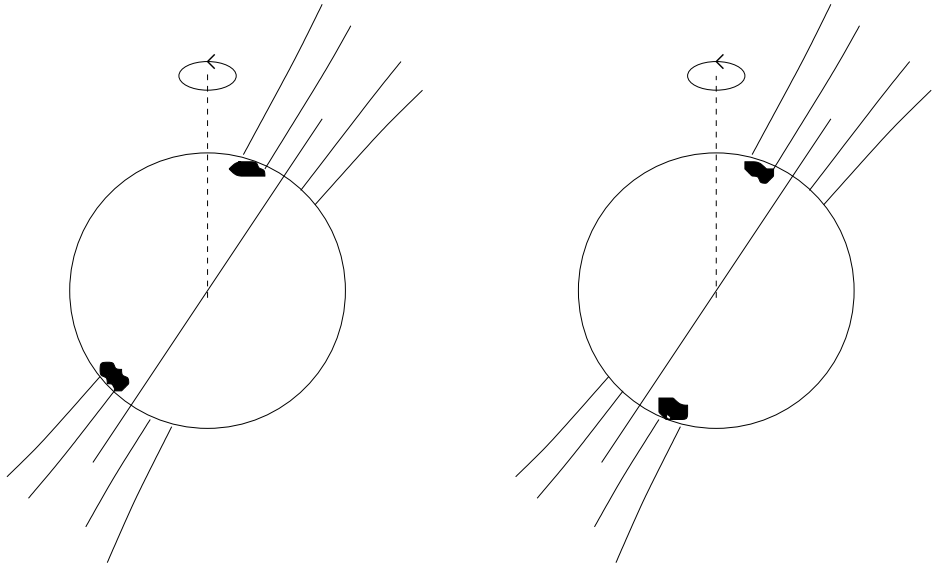}}
\end{picture}
\end{center}
\caption{Our two accretion scenarios. In the left hand panel we show
our first scenario in which the accretion flow is directed onto one or
other footprint of the same set of magnetic field lines at all
spin-orbit beat phases. The right hand panel shows our second scenario
in which the flow is directed onto a diametrically opposite set of
field lines at different beat phases. The spin axis of the white dwarf
is shown as a dotted vertical line.}
\label{dipole} 
\end{figure*}

We now examine two possible accretion scenarios: one in which the
accretion flow is directed onto one or other footprint of the same set
of magnetic field lines at all spin-orbit beat phases and the other in
which the flow is directed onto roughly diametrically opposite field
lines at different beat phases (see Fig \ref{dipole}). In the first
scenario we would expect the positive and negative circular
polarisation humps in the spin folded circular polarisation data (the
top panel of Fig. \ref{beat_phase}) to be seen at roughly similar spin
phases. In the second scenario we would expect to observe positive and
negative polarisation humps at distinct spin phases. The first
scenario is consistent with Fig. \ref{beat_phase}. 

We now consider the orbitally folded circular polarisation data. For
the first scenario we would expect the positive circular polarisation
hump to advance in phase as the white dwarf precesses. When the flow
is directed onto the other pole, the negative circular polarisation
hump will be roughly 180$^\circ$ out of phase with the positive
polarisation hump. In contrast, the second scenario would give
positive and negative humps at roughly the same orbital phase. Again
the first scenario is consistent with Fig. \ref{beat_phase}.

It is clear that the scenario in which the accretion flow is directed
onto roughly diametrically opposite points as we move in spin-orbit
beat phase is not consistent with the circular polarisation
data. Rather the data are consistent with the accretion flow being
directed onto the same field line at all spin-orbit beat phases. This
leads to the conclusion that at some spin-orbit beat phase the
accretion flow must follow a path around the white dwarf to accrete
onto the white dwarf rather than accrete onto the field lines in the
most direct possible way. This may be possible if one magnetic pole
was much stronger than the other: Schwope et al (1997) suggest that
one pole is strong in the EUV while the other is stronger in hard
X-rays which is consistent with this view. This would imply that there
is large dipole offset or equivalently that the magnetic field is more
complex than a dipole field. It is also possible that the magnetic
field is distorted perhaps as a result of the accretion
stream-magnetic field interaction.

As the accretion flow precesses around the white dwarf it will attach
onto different magnetic field lines and the accretion region on the
white dwarf will gradually shift, mainly in magnetic longitude, around
the magnetic axis of the white dwarf (Geckeler \& Staubert 1997). This
manifests itself in the spin-folded data: the peak of the positive
modulation advances in spin phase as we increase in beat phase
(Fig. \ref{beat_bin}). However, since the peak of the positive and
negative circular polarisation humps differ by only 0.2 spin cycles
this implies that the upper and lower accretion regions are fixed to
within $\sim70^{\circ}$ in magnetic longitude.  More detailed
modelling of the variations of the circular polarimetry will require
additional data, preferably simultaneously with X-ray observations.

\section{The near-synchronous polars}

The near-synchronous polars provide the best opportunity to
investigate the magnetic field structure of the white dwarf: we can
see directly the effect of the orientation of the magnetic field on
the way the accretion flow threads onto the field. This threading
process, which is not well understood, determines most of the
subsequent emission processes at the surface of the white dwarf in
both X-rays and the optical. In the near-synchronous systems, if we
can sample the spin-orbit beat period sufficiently, we are able (from
modelling the polarisation) to determine the accretion structures at
each orientation of the field, on a timescale in which other
parameters such as the mass transfer rate do not change very
significantly.

RX J2115--5840 is the fourth near-synchronous polar to be
discovered. (We note that Mukai 1998 proposes an alternative model for
V1432 Aql in which it is an intermediate polar with a spin period of
$\sim$67 min and an orbital period of 202 min). The first three such
systems (Table \ref{nonpolars}) had orbital periods which clustered
closely around 200 mins giving rise to some speculation that such an
orbital period was special in some way for these objects. The
discovery of RX J2115--5840 with an orbital period of 110 mins suggests
that it is not.

Of the four currently known near-synchronous polars, V 1432 Aql (RX
1940--10; Watson et al 1995, Friedrich et al 1996, Geckeler \&
Staubert 1997) and BY Cam (Silber et al 1997, Mason et al 1998) have
beat periods of weeks which makes it difficult to obtain sufficient
polarimetric coverage. V 1500 Cyg has a shorter beat period of 8 days
-- but the semi-amplitude of the circular polarisation is only 1.5\%
(Stockman, Schmidt \& Lamb 1988). We now have a system, RX J2115--5840,
which although relatively faint $V\sim$17--18 (a similar brightness to
V 1500 Cyg), has a beat period which is short enough, 6.3 days, to
obtain sufficient polarimetric coverage over the beat period. Such
coverage will allow us to determine, in principle, the magnetic field
structure of this system.

\begin{table}
\begin{tabular}{lllrr}
\hline
Star&$P_{s}$&$P_{o}$&$P_{o}-P_{s}/$&$P_{s}-P_{o}$\\
    & (mins)& (mins)&$P_{o}$ & (days)\\
\hline
V1432 Aql&202.507&201.940&-0.28\%&--49.5\\
BY Cam&199.330&201.258&0.96\%&14.5\\
V 1500 Cyg&197.502&201.043&1.79\%&7.8\\
RX J2115--58&109.547&110.889&1.21\%&6.3\\
\hline
\end{tabular}
\caption{The near-synchronous polars - the spin period of the white
dwarf, ($P_{s}$), the orbital period, ($P_{o}$) and the spin-orbit 
beat period ($P_{s}-P_{o}$).}
\label{nonpolars}
\end{table}

\section{Acknowledgments}

We would to thank the Director of SAAO, Dr R Stobie, for the generous
allocation of observing time and we are grateful to Dr D O'Donoghue
for the use of his period analysis software.

\end{document}